\documentclass[]{spie}  

\usepackage{amsmath,amsfonts,amssymb}
\usepackage{graphicx}
\usepackage{mathtools}

\title{Developing linear dark-field control for exoplanet direct imaging in the laboratory and on ground-based telescopes}
\author[a,b]{Thayne Currie}
\author[a,c]{Eugene Pluzhnik}
\author[a]{Ruslan Belikov}
\author[b]{Olivier Guyon}
\affil[a]{NASA-Ames Research Center, Moffett Field, California, USA}
\affil[b]{Subaru Telescope, National Astronomical Observatory of Japan, 650 N. Aohoku Pl., Hilo, Hawai'i, USA}
\affil[c]{Bay Area Environmental Research Institute, Oakland, CA, USA}

\authorinfo{Further author information: (Send correspondence to T. Currie)\\E-mail: currie@naoj.org}

\begin{document} 
\maketitle

\begin{abstract}
 Imaging rocky planets in reflected light, a key focus of future NASA missions and ELTs, requires advanced wavefront control to maintain a deep, temporally correlated null of stellar halo at just several diffraction beam widths.   We discuss development of Linear Dark Field Control (LDFC) to achieve this aim.  We describe efforts to test spatial LDFC in a laboratory setting for the first time, using the Ames Coronagraph Experiment (ACE) testbed.   Our preliminary results indicate that spatial LDFC is a promising method focal-plane wavefront control method capable of maintaining a static dark hole, at least at contrasts relevant for imaging mature planets with 30m-class telescopes.   
\end{abstract}


\section{INTRODUCTION}
\label{sec:intro}  
\begin{figure} [ht]
   \begin{center}
  \centering
   \includegraphics[scale=0.9,clip]{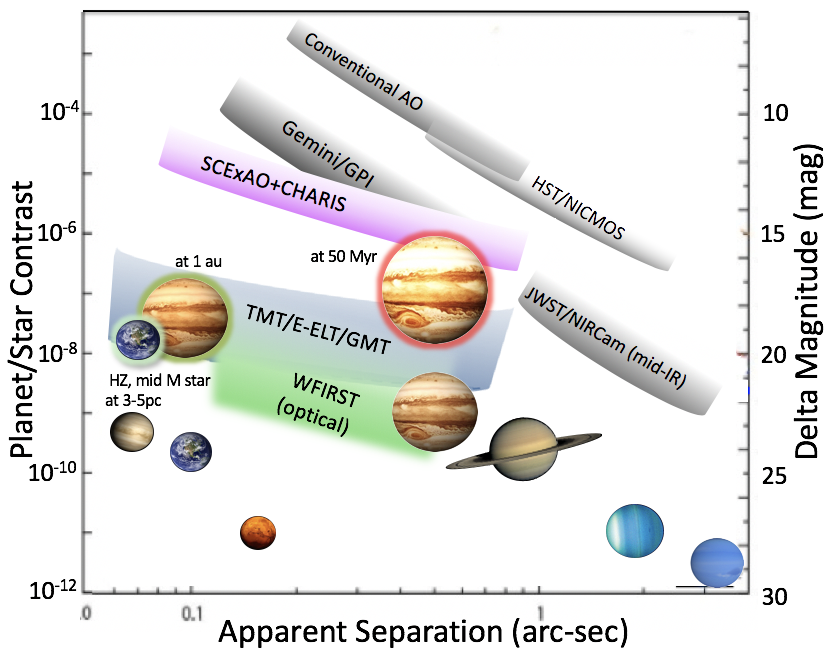}
   \end{center}
   \caption
   { \label{fig:contrast} 
    The demonstrated (expected) performance of past (upcoming) high-contrast imaging systems compared to that needed to image solar system analogues.   Contrasts are shown for mature solar system planets, a self-luminous young Jupiter (red glow), a Jupiter in reflected light at 1 au (green glow) for a system at 10 $pc$, and a habitable zone Earth-sized planet around a mid M star at 3--5 $pc$.  Contrasts for different objects are shown for 1.6 $\mu m$ observations around a Sun-like star unless otherwise noted.   Instrument system contrasts draw from multiple sources (e.g. \cite{Beichman2010,Brandt2014a,Brandt2014b,Guyon2018,Macintosh2014}): those listed for JWST/NIRCam refer to the `effective' 1.6 $\mu m$ contrast achieved for intermediate-aged stars with NIRCam at 4.4 $\mu m$ \cite{Skemer2014}.
}
   \end{figure}

Over the past decade, ground-based telescopes using facility adaptive optics (AO) systems and now dedicated \textit{extreme} AO systems have provided the first direct images of superjovian mass planets orbiting other stars (e.g.) \cite{Marois2008,Lagrange2010,Macintosh2015}.  Follow-up multi-wavelength photometry and spectroscopy\cite{Currie2011,Barman2015,Rajan2017} have yielded the first constraints on their atmospheric properties.
However, directly detecting and characterizing the spectrum of a habitable zone Earth-like planet around a Sun-like star with a future space mission requires suppression of noisy, scattered halo of starlight better by a factor of 10$^{10}$ (Figure \ref{fig:contrast}).  Planet-to-star contrasts required to image a  habitable-zone Earth around an M star are a factor of 100 milder ($\sim$ 10$^{-7}$--10$^{-8}$) but require detection at $\sim$ 20--100 milliarcseconds, typical $\approx$ 1--5$\lambda$/D at near-infrared wavelengths capable of revealing evidence for biomarkers (e.g. the 1.27 $\mu$m oxygen line\cite{LopezMorales2019}).  

New laboratory advances simulating space-based high-contrast imaging and new ground-based extreme AO systems have significantly advanced and demonstrated our ability to image fainter planets.   High-contrast imaging testbeds utilizing focal plane wavefront control (WFC) techniques like electric field conjugation (EFC; \cite{Give'on2007}) and advanced coronagraphy can generate a dark hole (DH) around a star.  Achieved null depths in monochromatic light and narrow bandpasses ($\lesssim$ 10$^{-8}$) are, \textit{if sustained}, sufficient to image reflected-light jovian planets, even around obscured apertures like WFIRST-CGI\cite{Seo2018,Shi2018}.

On ground-based telescopes, data obtained with extreme AO systems and processed with advanced algorithms yield near-infrared (near-IR) contrasts 10--100 times deeper than achieved with previous systems, imaging cooler ($T$ $\sim$ 700 $K$) young jovian planets with Saturn-like orbits \cite{Macintosh2015,Vigan2015}.  The newest extreme AO systems -- e.g. SCExAO and MagAO-X -- will soon achieve another factor of 10 contrast gain at small angles due to improved wavefront sensing and control  \cite{Jovanovic2015,Males2018,Currie2019}.   Successors to these extreme AO systems on upcoming \textit{extremely large telescopes} (ELTs) are being designed to reach raw contrasts of $\sim$ 10$^{-6}$ at small angles.  Should the stellar halo be further suppressed via post-processing at levels seen with current high-contrast imaging platforms, ELTs may provide the first direct detections of Earth-like exoplanets around nearby low-mass stars and probe evidence for biomarkers\cite{Guyon2018,LopezMorales2019}.

While high-contrast imaging testbeds simulating space-based high-contrast imaging have demonstrated significant progress towards the goal of imaging an Earth twin, maintaining a dark hole (DH) to see solar system-like planets requires extremely precise stellar halo measurements.   DH maintanence when the halo used for focal-plane wavefront control (FPWFC) itself is dark/low in flux (because FPWFC methods like EFC are applied in the first place) is problematic.   In scenarios, for example, with WFIRST-CGI, the dark hole will be photon starved.   By modulating the deformable mirror (DM) to determine and update the estimate of the electric field, FPWFC methods like EFC would perturb science exposures, potentially limiting exposure times.   In lieu of using the science target itself for FPWFC, another strategy is to first dig a DH around a much brighter reference star within 15-20$^{o}$ of a science target, apply the high-order deformable mirror correction achieving this DH to science observations \cite{Bailey2019}.   However, in this scenario the DH can still degrade -- both in terms of average intensity and temporal correlation with the initial DH at the start of science observations -- due to any number of dynamic aberrations.   The brightening of the DH and its decorrelation over time degrades the effectiveness of post-processing methods to remove residual starlight impeding planet detection.

   \begin{figure} [ht]
   \begin{center}
   \includegraphics[scale=0.65,clip]{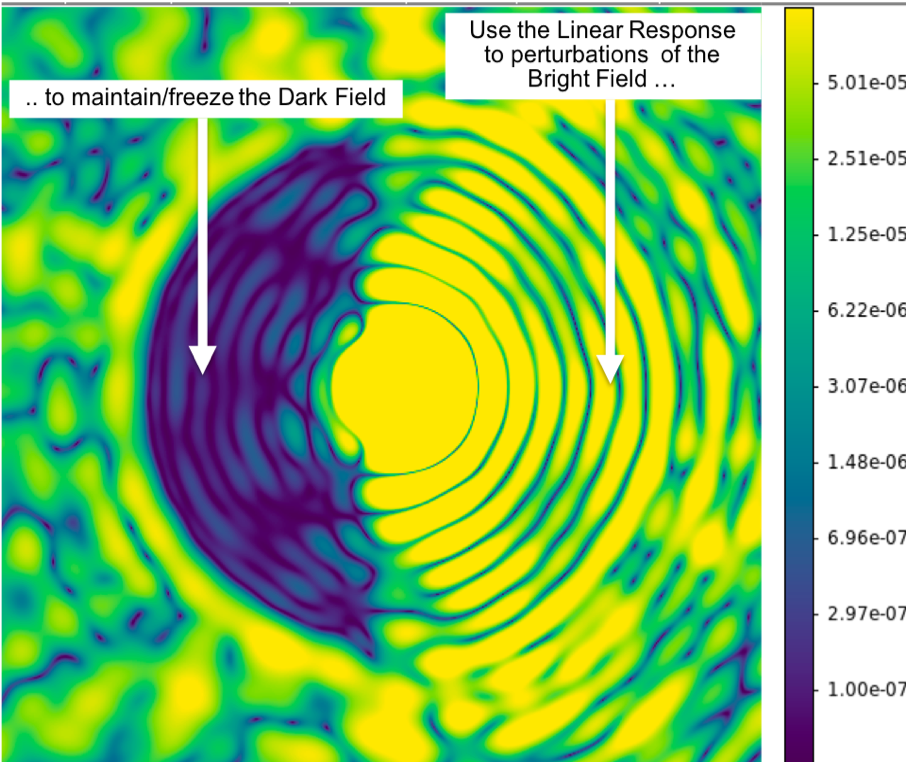}
   \end{center}
   \caption
   { \label{fig:ldfcschematic} 
   One-sided dark hole created from simulated Ames Coronagraphy Experiment data with labeling illustrating a schematic of Linear Dark Field Control.   In this example, regions in the bright field with an average contrast of $\sim$ 10$^{-4}$ are used to stabilize a dark hole with a contrast of $\sim$ 10$^{-7}$--10$^{-8}$.}
   \end{figure}  
   
Linear Dark Field Control\cite{Miller2017} (LDFC) is a promising wavefront control method which could maintain a static, deep DH that is first generated from FPWFC methods like EFC.   It utilizes the linear response of the uncorrected region in the focal plane which has far larger signal (the ``bright field" or BF) to wavefront perturbations that affect both the BF and the DF.   LDFC does not require modulating the signal within the DH and therefore requires only a single focal plane image to correct for the change in the electric field from the initial EFC-estimated state.    LDFC can be implemented in at least two ways: 1) ``Spatial" LDFC in a single band image, where a DH is created on one side of the image and stabilized by the BF on the opposite side\cite{Miller2017} as shown in Figure \ref{fig:ldfcschematic} or "Spectral" LDFC where the BF draws from pixels in image slices at wavelengths bracketing the bandpass within which the DH is created\cite{Guyon2017}.   Numerical simulations show that an LDFC control loop is able to hold static a DH at a contrast level comparable to that initially created using EFC, motivating laboratory tests to validate its efficacy at contrast levels needed for ground-based imaging of rocky planets with ELTs ($\sim$ 10$^{-5}$--10$^{-6}$ raw contrast) and later for WFIRST-CGI or NASA flagship missions like HabEx or LUVOIR ($\sim$ 10$^{-8}$--10$^{-10}$).  

In this work, we describe the first steps to validate and mature Spatial LDFC (hereafter "LDFC") in a laboratory setting using the Ames Coronagraph Experiment (ACE) testbed, at contrast levels relevant for direct imaging of mature exoplanets in reflected light with upcomig telescopes like the \textit{European Extremely Large Telescope} (E-ELT), \textit{Giant Magellan Telescope} (GMT), and \textit{Thirty Meter Telescope}.    

\section{Background}
\subsection{LDFC Theory}
Miller et al.\cite{Miller2017} describe in detail the theoretical premise of LDFC, which is based on the response of corrected, deep-contrast regions and uncorrected, shallow-contrast regions to the same telescope aberrations.   Briefly, the electrical field in image plane at a given time $t$ can be described as the sum of the incident pupil plane electric field $E_{\rm o}$ and a small change in complex amplitude induced by the deformable mirror, $E_{\rm DM}$:
\begin{equation}
\label{eq:efield}
|E_{\rm t}| \approx |E_{\rm o}| + |E_{\rm DM}|.
\end{equation}

The resulting intensity in the image plane, $I_{\rm t}$ = $|E_{\rm t}|^{2}$, is then comprised of three terms, which are due to the initial electric field, changes in the field due to the DM pertubation, and the inner product between the two:
\begin{equation}
\label{eq:i3terms}
|I_{\rm t}| \approx |E_{\rm o}|^{2} + |E_{\rm DM}|^{2} + 2<E_{\rm o},E_{\rm DM}>.
\end{equation}

 \begin{figure} [ht]
   \begin{center}
   \includegraphics[scale=0.425,clip]{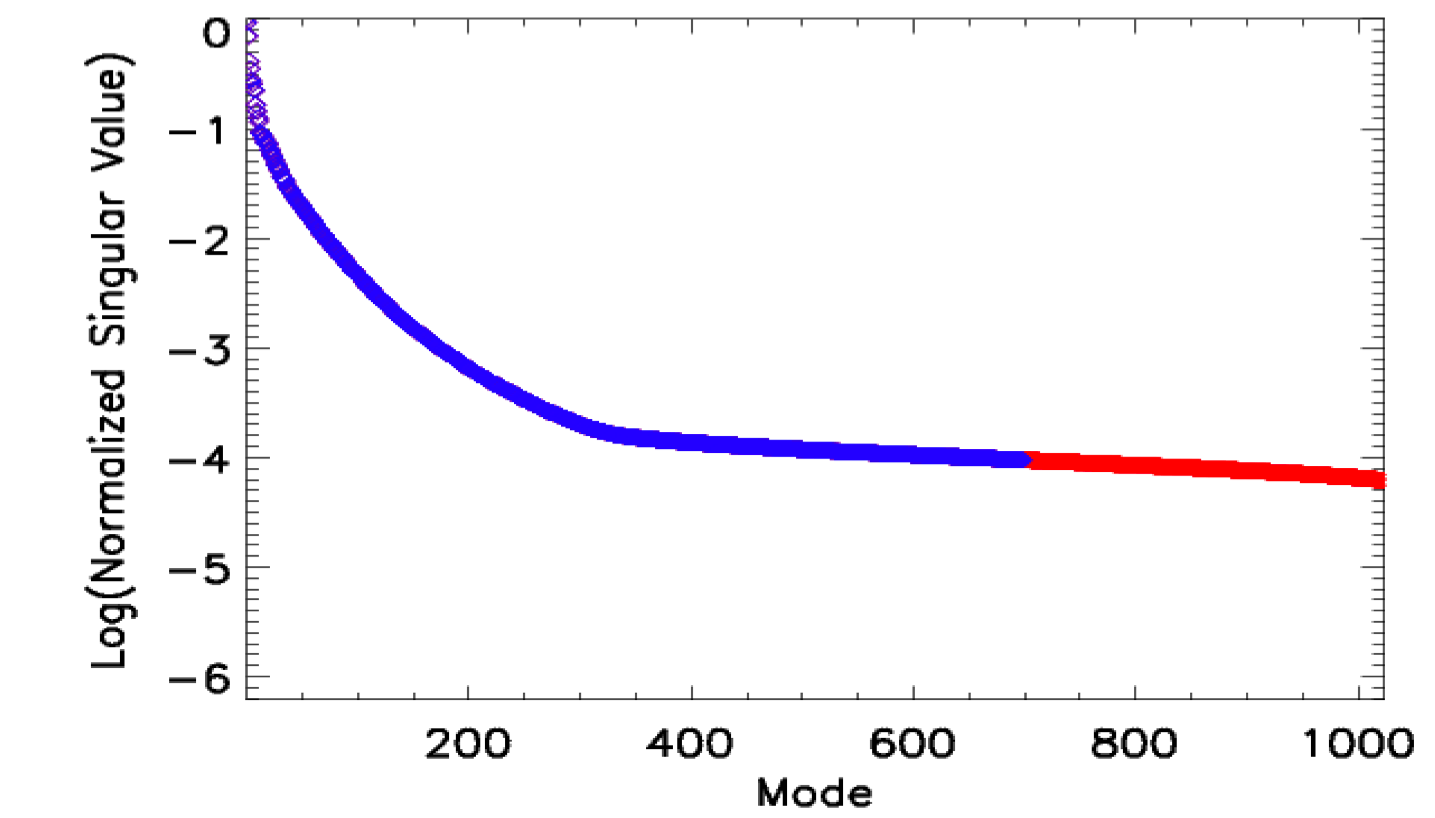}
   \includegraphics[scale=0.47,clip]{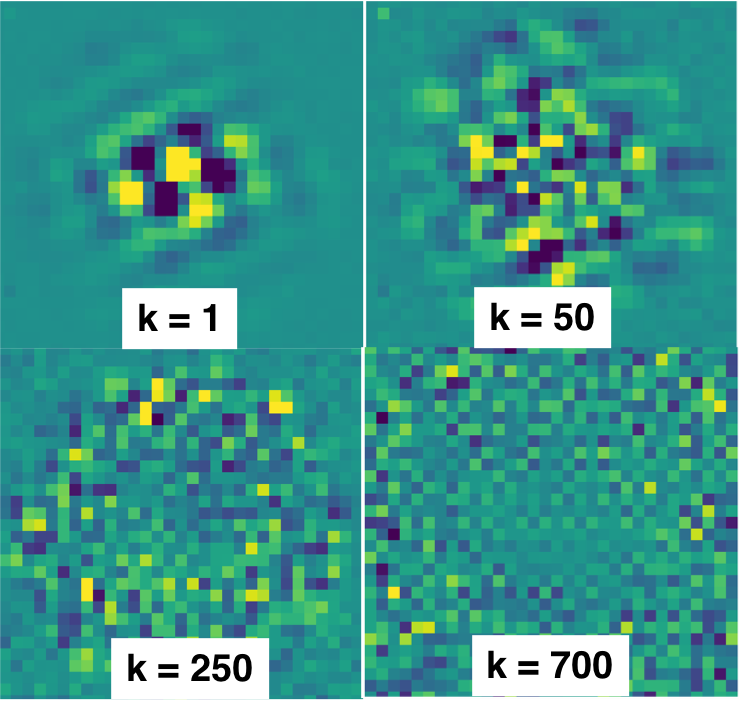}
   \end{center}
   \caption
   { \label{fig:svdlim} 
   Normalized singular values of the LDFC response matrix (left) and four modal DM responses (right).   In this case, we truncate the response matrix at 700 modes (modes shown in blue) and remove higher order modes (red).}
   \end{figure}

While $|E_{\rm DM}|^{2}$ dominates the intensity distribution within the dark field, the incident electric field is primarily responsible for the intensity distribution in the bright, uncorrected region, such that $|E_{\rm o}|^{2}$ $>>$ $|E_{\rm DM}|^{2}$ and $2<E_{\rm o},E_{\rm DM}>$ $>>$ $|E_{\rm DM}|^{2}$.   Therefore, the change in the intensity $I$ between time 0 and $t$ in the BF appears as a linear function of the change in complex amplitude induced by the DM perturbation: $\Delta$$I$ = $I_{\rm t}$ - $I_{\rm o}$  $\approx$ $2<E_{\rm o},E_{\rm DM}>$.  

By measuring changes in the bright field intensity at time $t_{\rm o}$ when the DH is established and time $t$ where it is corrupted, we can then determine the set of DM actuator offsets that restore the initial dark field.       An influence function -- e.g. the system response matrix, $RM$, with dimensions of $n$ pixels by $m$ actuators -- links together changes in DM shape $u$ to changes in the bright field intensity distribution: $\Delta$$I_{\rm DM, t}$ = $RM$$u_{\rm t}$.   Provided that bright field intensity recovery to its initial state simultaneously restores the dark field, actuator offsets $u$ required to drive the dark field back to its original state at time $t$ are then equal to the pseudo-inverse of $RM$ multiplied by $\Delta$$I_{\rm BF}$:
\begin{equation}
    u_{\rm t} = -(RM^{T}RM)^{-1}RM^{T}\Delta I_{\rm t, BF}.
\end{equation}

\subsection{Experimental Setup}

To provide an empirical test of LDFC, we used the Ames Coronagraph Experiment (ACE) laboratory at NASA-Ames Research Center.  The testbed experiments use a S1FC635 laser centered on 635nm as a monochromatic light source.  A PIAA coronagraph is used to suppress scattered starlight.    To achieve an initial flat wavefront at the pupil plane, we use an implementation of the Gerchberg-Saxton method, using a sequence of random pupil plane phase probes\cite{Pluzhnik2017}.  The D-shaped focal plane occulter normally blocking all of one side of the wavefront sensor image was removed, making visible the full 360 degree field needed for LDFC.

Using a speckle nulling wavefront control loop as implemented in previous ACE testbed experiments (e.g. \cite{Belikov2012}), we created a one sided, C-shaped dark hole extending between 1.2 $\lambda$/D and 4.5 $\lambda$/D.   Due to our removal of the occulter, internally reflected light partially contaminates three isolated regions of the dark hole.   The average dark hole intensity contrast with respect to the PSF core is roughly $\sim$ 10$^{-5}$: a contrast below that currently achieved at 1--5 $\lambda$/D with extreme AO systems on 8-10m telescopes but approaching that required for imaging mature planets in reflected light around M stars with ELTs.   At 1.2--4.5 $\lambda/D$, the average intensity in the bright, uncorrected region ranges between 10$^{-3}$ and 10$^{-4}$.



We calculate the spatial LDFC response matrix ($RM$) by perturbing each of the $m$ actuators by a pair of small amplitude pokes, $ampl_{\rm poke}$,
in the positive and negative direction and recording the intensity $I$ over $n$ BF pixels.   We combine results from two separate poke patterns -- $a$ and $b$ -- which differ in the sign (up or down) of a given actuator poke:
\begin{equation}
    RM(n,m) = 0.5*[(I_{\rm a_{\rm 1}}-I_{\rm _{\rm a2}})+(I_{\rm b_{\rm 1}}-I_{\rm b_{\rm 2}})]/(2*ampl_{\rm poke})
\end{equation}

The control matrix ($CM$) in a closed-loop implementation of LDFC is the pseudo-inverse of $RM$: 
\begin{equation}
    CM = (RM^{T}RM)^{-1}RM^{T}
\end{equation}.    

To compute $CM$, we decompose $RM^{T}RM^{-1}$ into eigenvectors $V$ and eigenvalues $\Lambda$ and truncate $\Lambda$ at mode $k_{lim}$ (Figure \ref{fig:svdlim}a) before inverting to yield the $CM$: 
$CM$ = ($V$${\Lambda}^{-1}$$V^{T}$)$_{\rm k<k_{lim}}$$RM^{T}$. 
For our set up, the normalized singular values of the $RM$ covariance flatten to 10$^{-4}$ between $k$ = 250 and $k$ = 1024 (where the $RM$ covariance would be at full-rank).   We set the modal cutoff to an intermediate value consistent with the highest mode where the imprint of a circular region of the DM (excluding the corners) is still visible (Figure \ref{fig:svdlim}b).

To test the efficacy of spatial LDFC, we stored the DM shape producing our dark hole and then introduced a grid of DM perturbations to degrade it.   Each perturbation corresponds to a single actuator poke -- as in our $RM$ calculations -- instead of a sine wave modulation producing a pair of speckles in the BF and DF or a Kolmogorov phase screen.   We set the perturbation amplitude such that it produced a factor of 2--5 degradation in the average DF intensity, intermediate between the original DF intensity and the BF intensity.   

\begin{figure} [ht]
   \begin{center}
   \includegraphics[scale=0.7,clip]{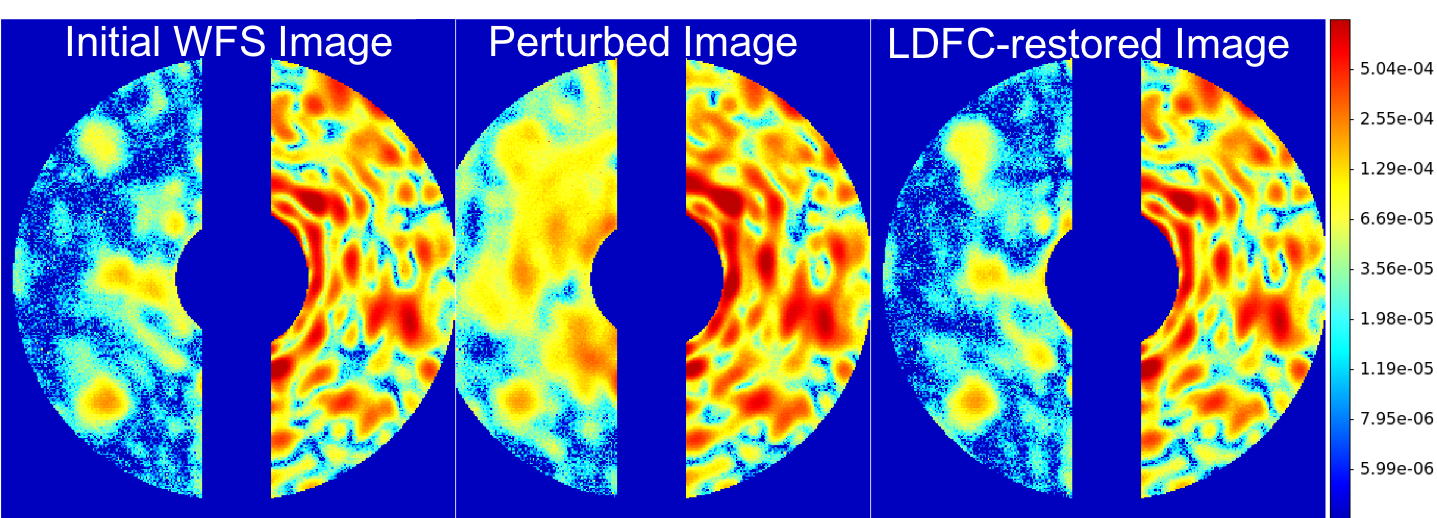}
   \end{center}
   \caption
   { \label{fig:ldfcsequence} 
   A demonstration of spatial LDFC.   We introduced a single actuator DM perturbation onto a flat WFS image, degrading the DH by a factor of $\sim$ 3.   After 20 iterations with a gain of 0.1, our closed-loop LDFC implementation largely recovers the flat WFS state, yielding a nearly identical average intensity. }
   \end{figure}  

For each perturbation, an iterative closed-loop implementation of spatial LDFC determined actuator offsets that should drive the BF (and, in turn, the DF) back to its original average intensity.
The average WFS intensity within the DH and variance in the residuals between it and the original DH were tracked over 20 iterations for each perturbation.  We varied the gain $g$ to identify values that led to convergence in most cases after 20 iterations.   Exposure times are tuned to fully illuminate the bright field without saturation pixels and illuminate the dark field where photon shot noise is slightly below our nominal (speckle limited) contrast of 10$^{-5}$.

\section{Preliminary Results}
Our preliminary results demonstrate the promise of spatial LDFC to stabilize the average dark hole intensity and sustain a high temporal correlation of the dark field at contrast levels relevant for future ground-based high-contrast imaging of planets with ELTs in reflected light.  Figure \ref{fig:ldfcsequence} shows a successful convergence of the LDFC loop.   The initial dark hole with a contrast of $\sim$ 10$^{-5}$ is degraded by a single-actuator perturbation which increases the average intensity by a factor of 3.     After 20 iterations, the average intensity of both the bright field and dark field have returned to their initial values.   The expected change in DM shape in this case -- a single actuator poke whose sign is opposite the initial perturbation -- is recovered.

\begin{figure} [ht]
   \begin{center}
   \includegraphics[scale=0.7,clip]{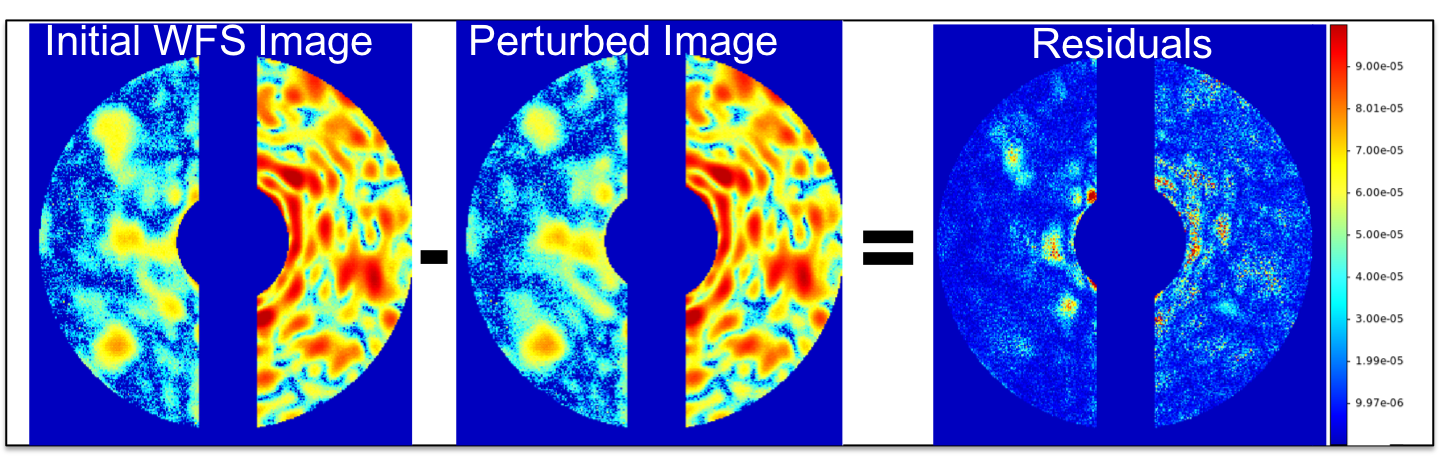}
   \end{center}
   \caption
   { \label{fig:ldfcresiduals} 
   Residuals of the difference between the initial wavefront sensor image with a 10$^{-5}$ dark hole and that restored using spatial LDFC.   Three of the bright residual regions correspond to static, uncorrectable features due to system internal reflection.}
   \end{figure}

In the example shown in Figure \ref{fig:ldfcsequence}, the temporal correlation of the dark hole has also been restored.   Subtraction residuals between the initial and perturbed dark hole are about a factor of 2.5 larger than the average intensity of the initial dark hole itself.   After 20 iterations, the subtraction residuals in the dark hole have dropped to the noise limit of $\sim$ 7$\times$10$^{-6}$ contrast (Figure \ref{fig:ldfcresiduals}).

     \begin{figure} [ht]
   \begin{center}
   \includegraphics[scale=0.45,clip]{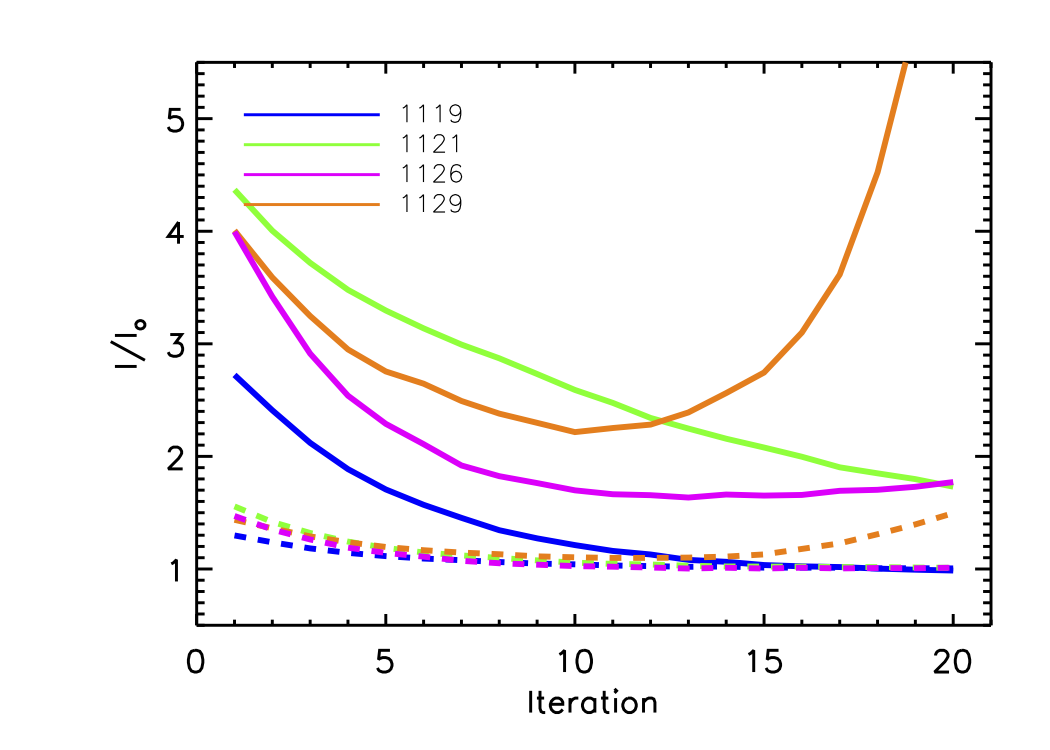}
   \includegraphics[scale=0.45,clip]{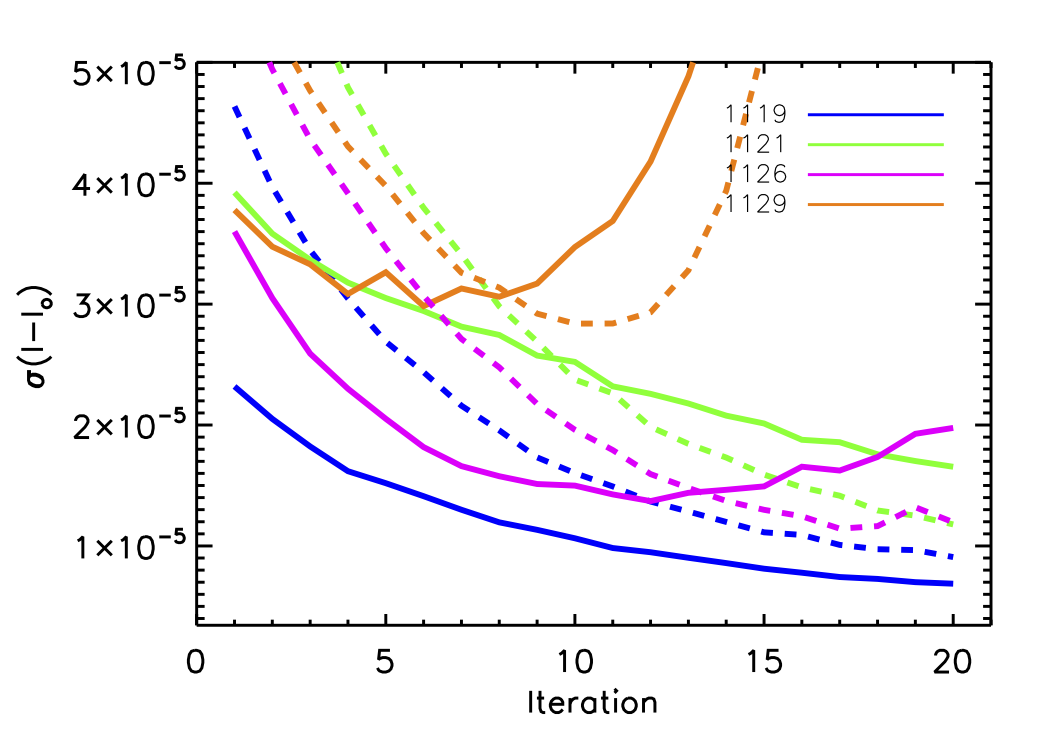}
   \end{center}
   \caption
   { \label{fig:outcomes} 
   Contrast (with respect to the initial dark hole) and subtraction residuals vs. iteration for four different perturbations corrected by our spatial LDFC loop.   Dashed lines refer to the bright field; solid lines show results for the dark field.  The numbers refer to the poke index, which has dimensions of 1024 actuators $\times$ 2 = 2048.   The perturbations that are shown applied a change to the shape of the deformable mirror near its center.}
   \end{figure} 
   
   Other cases show a range of outcomes, which are depicted in Figure \ref{fig:outcomes}.   In many cases, spatial LDFC successfully drive the dark hole back to its initial state after 20 iterations (loop number 1119).   For some DM perturbations whose response is weak in the focal plane, the dark field is degraded only by $\sim$ 10-20\%.  For others, the loop is still converging after 20 iterations (loop number 1121), or it has stalled out at a value a factor of $\sim$ 1.5--2 above the initial dark hole contrast (loop number 1126).   Finally, in some cases, the spatial LDFC loop initially drives the dark field back to its original state but then destabalizes after 5-10 iterations (loop number 1129).

\section{Discussion}
Spatial Linear Dark Field Control is novel method for mainting the average intensity and temporal correlation of a dark hole initially created through FPWFC methods.   \cite{Miller2017} demonstrated the promise of this technique through sophisticated numerical simulations showing that spatial LDFC can cancel speckles introduced into the dark field at a deep contrast level ($\sim$ 10$^{-7}$--10$^{-8}$) within a small number of iterations.  Our preliminary laboratory tests show evidence for the viability of spatial LDFC in real systems, as our closed-loop implementation typically restores (much of) a dark hole at a contrast of 10$^{-5}$ within 20 iterations.   

Key challenges for demonstrating spatial LDFC focus both on its practical implementation and on its fundamental limits.   For some perturbations, the spatial LDFC loop goes unstable after about 5-10 iterations.   Examination of the actuator offsets predicted by spatial LDFC show that in these cases the expected change in DM shape is not recovered.  Better regularization of the CM and the closed-loop correction as well as modal weighting of the CM may allow the spatial LDFC loop to converge to a sustained, correct DM shape. 

Alternatively, some perturbations may be sufficiently strong that throw the bright field response out of the regime where a linear approximation is valid: unstable loops almost always occur where the initial perturbed dark field is a factor of $\sim$ 5 or more brighter than the unperturbed field.   It is also possible that these perturbations may be affected by uncorrectable regions within the bright and dark fields, which should be masked prior to closed-loop implementation of spatial LDFC\footnote{While a null space between the bright field and dark field is similarly possible, perturbations evincing such a case should return the bright field back to its original state but leave the dark field unaffected, which has not been seen so far.}.    

Spatial LDFC in contrast regimes tested here should be directly applicable to ground-based imaging of planets in reflected light with ELTs \cite{Guyon2018}.   For real systems and at slightly deeper dark hole contrasts ($\sim$ 10$^{-6}$), the dynamic range required to sense bright field perturbations and illuminate the residual halo within the dark field may nominally require an enormous dynamic range, although this could be in part compensated by a partially transmissive mask over the bright field.  For NASA missions like WFIRST CGI where the dark hole subtends a full 360 degrees and its contrast is $\sim$ 10$^{-4}$ times brighter than an uncorrected region, other solutions may be required, such as splitting the incident light into two separate, corrected images with one-sided dark holes before recombination or utilizing \textit{spectral} LDFC \cite{Guyon2017}.

Near-term tests of spatial LDFC will focus on comparing its performance to speckle nulling and EFC both the high flux regime (where the dark hole is well illuminated) and the low-flux regime (where it is not).  Additional testing LDFC within the ACE lab environment includes (a) using pupil masks representing future NASA missions, such as HabEx and LUVOIR; (b) vacuum tests; (c) on-sky testing with the Subaru Coronagraphic Extreme Adaptive Optics project at the Subaru Telescope in Hawai’i.  The former will test LDFC at contrast levels more relevant for Earth-imaging.   The latter will provide a test of LDFC at leading, deep contrasts on the ground within the context of an integrated system.   Both will help push the technological readiness level for LDFC.
 

\acknowledgments 
All coauthors are supported by a NASA Strategic Astrophysics Technology award ``Linear Wavefront Control for High-Contrast Imaging".   T. C. is also supported by a NASA Senior Postdoctoral Fellowship.   We thank Mengshu Xu for help with preparation of Figure 1 in this publication.


\bibliography{report} 
\bibliographystyle{spiebib} 

\end{document}